\documentclass[aps,prl,reprint,amsmath,amssymb,preprintnumbers,10pt,floatfix,nofootinbib]{revtex4-2}
\usepackage{amsmath}
\usepackage{graphicx}
\usepackage{subfigure}
\usepackage{amssymb}
\usepackage{xcolor}
\usepackage{multirow}
\usepackage{cancel}
\usepackage{color}
\usepackage{cancel}
\usepackage{ulem}
\usepackage{listings}
\usepackage{xcolor}
\usepackage{float}
\usepackage{slashed}
\usepackage{wrapfig}
\usepackage{mathtools}
\usepackage[colorlinks,citecolor=blue]{hyperref}
\usepackage[T1]{fontenc}  
\usepackage[utf8]{inputenc}  
\usepackage{times}
\usepackage{bbm}

\newcommand{\StartEndMatter}{%
  \setcounter{equation}{0}%
  \renewcommand{\theequation}{E\arabic{equation}}%
  \section*{End Matter}%
}

\DeclareMathOperator{\real}{Re}
\DeclareMathOperator{\imag}{Im}
\DeclareMathOperator{\Tr}{Tr}
\DeclareMathOperator{\diag}{diag}
\def\i{\text{i}}
\def\d{\text{d}}
\def\k{\boldsymbol{k}}

\begin{document}

\title{Plasma Effects Suppress Mixing-Induced Collisional Freeze-In}

\author{Shao-Ping Li}
\email{Shaoping.Li@oeaw.ac.at}
\affiliation{Marietta Blau Institute for Particle Physics, Austrian Academy of Sciences, Dominikanerbastei 16, A-1010 Vienna, Austria}

\author{Josef Pradler}
\affiliation{Marietta Blau Institute for Particle Physics, Austrian Academy of Sciences, Dominikanerbastei 16, A-1010 Vienna, Austria}
\affiliation{University of Vienna, Faculty of Physics, Boltzmanngasse 5, A-1090 Vienna, Austria}

\preprint{UWThPh 2026-12}

\begin{abstract}
Dark matter that is weakly mixed with particles in a thermal plasma
can be produced through both, oscillatory and collisional freeze-in. While plasma effects have long been known to suppress the former, their role in collisional freeze-in remains less clear quantitatively, leading to %
conflicting results in
sterile neutrino dark matter production via  collisional freeze-in. Using first-principles nonequilibrium Kadanoff-Baym equations in the flavour-covariant formulation, we establish the general result that plasma effects also suppress collisional freeze-in: the vacuum mixing angle in the collision rate is replaced by its plasma-corrected counterpart.
Applying this result, we conclude that sterile neutrinos cannot be dark matter in the keV-MeV mass range when produced from the Standard Model plasma.
\end{abstract}

\maketitle

\paragraph{\textbf{Introduction.}}
Dark matter (DM) production through freeze-in~\cite{McDonald:2001vt,Hall:2009bx} is a simple mechanism to explain the observed relic density today~\cite{Planck:2018vyg}. When there is mixing between the DM state and some thermal particle state,  DM production  from oscillations can also contribute to the relic density, which has been widely considered in sterile-neutrino DM via active-sterile neutrino mixing~\cite{Dodelson:1993je,Dolgov:2000ew,Abazajian:2001nj,Asaka:2006nq,Canetti:2012kh,Drewes:2016upu,Boyarsky:2018tvu}, and in dark photon DM production via photon-dark photon mixing~\cite{Pospelov:2008jk,Redondo:2008ec,Jaeckel:2008fi,An:2013yfc,Fradette:2014sza,Caputo:2021eaa}. Nevertheless, the oscillatory freeze-in  becomes strongly suppressed  due to plasma effects unless the production exhibits a resonance~\cite{Shi:1998km} similar to the  Mikheyev-Smirnov-Wolfenstein effect~\cite{Wolfenstein:1977ue,Mikheev:1986wj} in neutrino oscillations.

At higher temperatures, where oscillation production is typically inefficient, collisional freeze-in can dominate.
This possibility has received renewed attention in models of sterile neutrino DM~\cite{Kusenko:2006rh,Petraki:2007gq,Wu:2009yr,Besak:2012qm,Merle:2013wta,Lello:2016rvl,Coy:2021sse,Datta:2021elq,Li:2022bpp,Abada:2023mib}. 
Based on semiclassical Boltzmann kinetics, Ref.~\cite{Datta:2021elq} argued that decays of electroweak gauge bosons can produce the observed sterile-neutrino abundance for masses in the keV to MeV range, while permitting active-sterile mixing angles small enough to evade stringent bounds from diffuse X-ray and gamma-ray observations~\cite{Abazajian:2001vt,Roach:2019ctw,Foster:2021ngm,Krivonos:2024yvm}. The lower part of this mass range is further constrained from structure-formation~\cite{Boyarsky:2008xj,Irsic:2017ixq,Ballesteros:2020adh,Dekker:2021scf}, with the precise bounds depending on the momentum distribution generated by the production mechanism. However, an earlier study, which identified the neutrino damping rates in finite-temperature Dirac equations,  found that sterile-neutrino production from gauge interactions is suppressed by medium effects~\cite{Lello:2016rvl}, leading to apparently conflicting conclusions about the viability of sterile-neutrino DM in this mass range. 

Unlike in oscillatory freeze-in, plasma effects in collisional freeze-in are still not fully understood. Oscillatory production is naturally formulated in the interaction basis using the density-matrix formalism~\cite{Sigl:1993ctk}, since the production depends on the coherent evolution of interacting and weakly coupled states. 
Collisional production is usually described in terms of free momentum-eigenstates in Feynman-diagramatic calculations of individual reactions. Conventional Boltzmann treatments therefore often
use the vacuum mixing angle, with plasma effects entering at most through thermal masses in the kinematics~\cite{Coy:2021sse,Datta:2021elq,Li:2022bpp}.

In a plasma, however, the propagating eigenstates do not coincide with the vacuum-mass eigenstates. Adding thermal masses to vacuum dispersion relations accounts for changes in kinematics, but does not in general determine the medium-induced rotation of the propagating states. As a consequence, it also does not determine the overlap with the interaction state that couples a feebly interacting particle to the thermal bath. It is this overlap that controls the mixing-induced collision rates. A first-principles treatment of collisional freeze-in must therefore first identify the in-medium propagation eigenstates and then evaluate their production through the plasma interactions.

\begin{figure}[t]
	\centering
\includegraphics[width=\columnwidth]{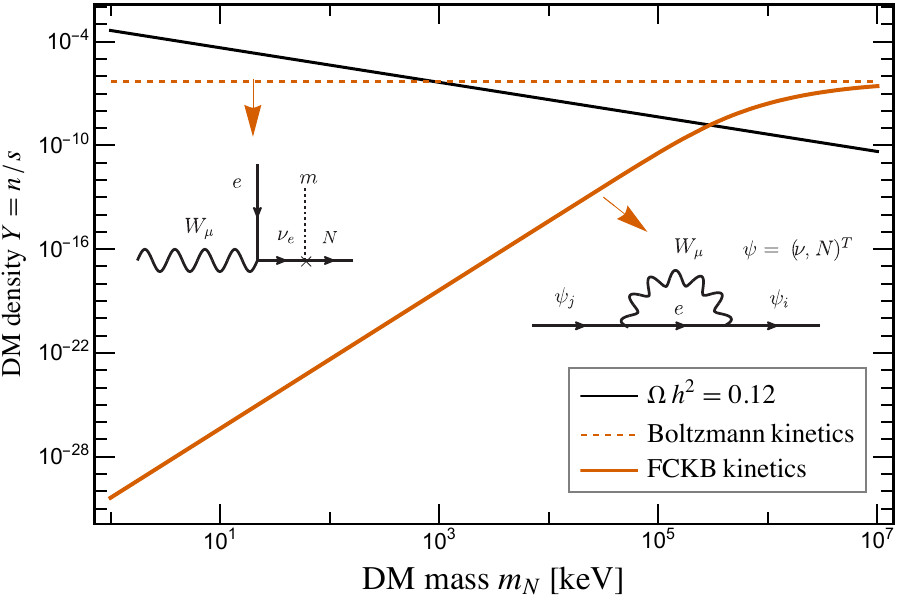}
    \caption{\label{fig:YmN} A typical example of mass-mixing induced collisional freeze-in production of a relativistic gauge singlet fermion $N$ from $W$ gauge boson decay. The black line shows the DM relic density yield $Y\equiv n/s$ that matches the observed value today, with $n$ the DM number density and $s$ the entropy density of the plasma. The dashed line is derived from the semiclassical Boltzmann kinetic equation, 
    induced by a mass mixing parameter $m$ between the left-handed SM neutrino $\nu_e$ and the right-handed DM $N$. The thick line is derived in this work under the  FCKB constraint and kinetic equations, by forming the  neutrino and DM eigenstates in a flavour-covariant $\psi=(\nu, N)^T$ basis.}
\end{figure}

In this \textit{Letter}, we determine these ingredients by using the \textit{constraint} and \textit{kinetic} equations of the flavour-covariant Kadanoff--Baym (FCKB) formulation~\cite{Prokopec:2003pj,Prokopec:2004ic,Beneke:2010dz,BhupalDev:2014pfm,Li:2026qym}. The constraint equation determines the in-medium propagation eigenstates and their relation to the interaction basis. The kinetic equation then describes the collisional production of these states. We thereby show that the resulting collisional freeze-in rate is governed by the plasma-corrected mixing angle, rather than by its vacuum value. 
Applied to sterile-neutrino DM, we show that this suppresses production from electroweak gauge-boson decays by several orders of magnitude relative to the semiclassical Boltzmann result, as illustrated in Fig.~\ref{fig:YmN}. This result confirms and generalizes the suppression found in~\cite{Lello:2016rvl}, resolves its conflict with~\cite{Datta:2021elq}, and conclusively settles the viability of keV-MeV sterile neutrino~DM. The identical plasma suppression effect between oscillatory and collisional freeze-in established in this Letter, together with recovering the semiclassical Boltzmann equation was, to the best of our knowledge, not yet explicitly established in quantum kinetic equations. Natural units $\hbar=c=k_B=1$ are used.

\paragraph{\textbf{Effective mixing angle.}}
To develop the FCKB equations for mixing-induced collisional freeze-in, we take mass-mixing for illustration. As a typical example, we consider a fermion gauge singlet $N$ as the DM particle coupling to the Standard Model (SM) electron neutrino~$\nu_e$. The plasma couples to~$\nu_e$ only, through interactions of the generic form $\mathcal{L}=
\mathcal{\bar O}_{\rm SM}P_L \nu_e+\text{h.c.}$, where $P_L$ reflects the left-handed nature of weak interactions and   $\bar O_{\rm SM}$ is built from thermalized SM fields. For the charged and neutral weak current $\bar O_{\rm SM}=-g_2\bar e \gamma^\mu W_\mu/\sqrt{2}$ and 
$\bar O_{\rm SM}=
-g_2 \bar \nu_e \gamma^\mu Z_\mu/(2\cos\theta_W)$, respectively, 
with $g_2$  the  $SU(2)_L$ gauge  coupling and $e$ the electron state and $\theta_W$ the weak mixing angle. In terms of the mass eigenstates $\psi_i$, $\nu_e = \sum_i U_{ei} \psi_i$, so that the mixing matrix $U$ induces an interaction of the vacuum eigenstate $N_m$ with the SM particles. 
Generalization to three neutrino flavours is straightforward, but here we assume that~$N$ mixes dominantly with~$\nu_e$. In a more general setup, one can consider all possible mixing portals between a thermal plasma and an out-of-equilibrium system, such as scalar~\cite{McDonald:1993ex,Fradette:2018hhl} and vector mixing~\cite{Holdom:1985ag}. The general conclusions drawn below do not depend on the specific structure of the operator $\mathcal{\bar O}_{\rm SM}$, provided that the fields furnishing $\mathcal{\bar O}_{\rm SM}$ are in thermal equilibrium.

Three bases enter the problem: the weak-interaction (flavour) basis $\psi^{\rm int}=(\nu_e,N)^T$ in which the weak gauge-interactions are diagonal, $N$ being a singlet, and which identifies the thermalized states in the plasma; 
the vacuum-mass basis $\psi^{\rm vac}=(\nu_m,N_m)^T$, and the in-medium (plasma) propagation basis~$\psi^{\rm pla}$.
Collision terms describe the production of propagating states through plasma interactions. The relevant coupling that controls mixing-induced production is therefore set by the rotation of the propagating states onto the interaction basis. We write the two required unitary basis transformations as
\begin{align}
\label{eq:bases}
\psi^{\rm vac}= U^\dagger \psi^{\rm int}, \quad \psi^{\rm pla}= U_{\rm eff}^\dagger \psi^{\rm int} ,
\end{align}
with $U^\dagger U=UU^\dagger=1$ and likewise for $U_{\rm eff}$. Here,~$U$ is the familiar vacuum mixing matrix and $U_{\rm eff}$ is the effective unitary mixing matrix entering the plasma collision term and which we must determine.
For the simple $2\times2$ mixing case, $U$ is a real rotation matrix with vacuum mixing angle $\vartheta\ll 1$, $U_{e1} = \cos\vartheta$, $U_{e2} = \sin\vartheta$,
and $U_{\rm eff}$ is parameterized likewise, with $\vartheta_{\rm eff}$ the effective mixing angle;
in vacuum $U_{\rm eff}\to U$.

The relations given in Eq.~\eqref{eq:bases} can be determined by the Kadanoff–Baym constraint equations, which identify the pole structure and flavour composition of the propagating states in the  medium. In the flavour-covariant formulation, we employ the approximate equation~\cite{Prokopec:2003pj,Prokopec:2004ic,Li:2026qym},
\begin{align}\label{eq:constraint}
   &2k_0\gamma^0 \i S^{<(>)}(k,t)
   \nonumber \\
   &=\left\{ \left({\bf k}\cdot\boldsymbol{ \gamma}+\mathbb{ M}+\real\Sigma^R(k)\right)\gamma^0, \gamma^0\i S_{}^{<(>)}(k,t)\right\},
\end{align}
where $\i S_{ij}^<(k,t)$ and $\i S_{ij}^>(k,t)$ are the Wightman two-point correlation functions obtained from $\i S^<_{ij}(x,y)\equiv-\langle\bar\psi_j(y)\psi_i(x)\rangle$ and $\i S^>_{ij}(x,y)\equiv \langle  \psi_i(x) \bar\psi_j(y)\rangle$, respectively, after a Wigner transformation in a spatially homogeneous background, with the relative Fourier four-momentum $k$, three-momentum ${\bf k}$, and average time $t=(x^0+y^0)/2$; $\{\cdot,\cdot\}$ denotes the anticommutator, and $\gamma^0,\boldsymbol{\gamma}$ are Dirac matrices. In Eq.~\eqref{eq:constraint}, $\mathbb{M}$ is a flavour-covariant mass matrix. In the vacuum-mass basis $\mathbb{M} = \mathbb{\hat M} = \diag(\hat m_1,\hat m_2)$ with $\hat m_1=m_\nu\approx 0$ and $\hat m_2=m_N$.
In  vacuum, one takes the  real part of the thermally retarded self-energy amplitude to zero, $\real\Sigma^R(k)=0$, from which one can derive the  quasiparticle spectrum for the free Wightman function with $\mathbb{M}=\mathbb{\hat M}$,
$\i S^{<,\rm vac}_{ij}\propto \,(\slashed{k}+\hat m_i \delta_{ij})\,\delta[k^2-(\hat m^2_i+\hat m_j^2)/2]$~\cite{KadanoffBaymCRC,Li:2026qym}. 

In the vacuum mass basis, $\mathbb{\hat M}$ is diagonal, but the thermal correction is not: the plasma couples to $\nu_e$ only, $\real\Sigma^R_{ij}=U^*_{ei}U_{ej}\real\Sigma^R_\nu$, with $\real\Sigma^R_\nu$ being the self-energy of the electron neutrino. For the propagating left-handed components the constraint equation~\eqref{eq:constraint} reduces to (see End Matter)
\begin{align}\label{eq:pole-eq}
\big[k^2-(\mathbb{\hat M}^2+\Pi)\big]\,\i S^<\simeq0\,,\qquad \Pi_{ij}=U^*_{ei}U_{ej}\,\Pi_\nu\,,
\end{align} 
where $\Pi_\nu\equiv\Tr_{\rm s}\big[\slashed k \real\Sigma^R_\nu\big]$ is the Dirac trace of the thermal correction of the electron neutrino, containing both charged and neutral weak currents. The plasma propagating states are therefore the eigenvectors of $\mathbb{\hat M}^2+\Pi$ and a further rotation $\psi^{\rm pla}=V\psi^{\rm vac}$, $\i S^{<,\rm pla}=V\,\i S^{<,\rm vac}V^\dagger$, is required,
\begin{align}\label{eq:diagonalization}
V\left(\mathbb{\hat M}^2+\Pi\right)V^\dagger=\diag(\tilde m_1^2,\tilde m_2^2)\,,
\end{align}
with $\tilde m_i^2$ the in-medium poles containing the vacuum masses and thermal-mass corrections. In the $\psi^{\rm pla}$ basis, the solution of Eq.~\eqref{eq:constraint} is
$\i S_{ij}^{<,\rm pla}\propto \slashed{k} \delta(k^2-[\tilde m_i^2+\tilde m_j^2)/2]$ in the relativistic limit. 

From Eq.~\eqref{eq:bases} we obtain $U_{\rm eff} = U V^\dag $. In the interaction basis $\Pi$ has the single non-zero entry $U\Pi U^\dagger=\diag(\Pi_\nu,0)$, and diagonalizing $U(\mathbb{\hat M}^2+\Pi)U^\dagger$ by $U_{\rm eff}$ (see End Matter) determines the effective mixing angle
\begin{align}\label{eq:theta_eff}
  \sin^2({2\vartheta_{\rm eff}})&=\frac{m_N^4 \sin^2({2\vartheta})}{\left(m_N^2 \cos({2\vartheta})-\Pi_\nu\right)^2+  m_N^4 \sin^2({2\vartheta})}\,.
\end{align}
Remarkably, the same effective  angle also appears in oscillatory freeze-in production via diagonalization of the effective Hamiltonian for oscillations~\cite{Dodelson:1993je}.
Here, instead, Eq.~\eqref{eq:theta_eff} is obtained from the FCKB constraint equation and is tailored to the collisional freeze-in production rate. 
It implies that the traditional Boltzmann result will be recovered in the asymptotic limit $m_N^2\gg |\Pi_\nu|$, which, however, cannot be obtained by simply adding the thermal-mass effects to the dispersion relation of propagating sates, as we will see below.
When thermal corrections dominate, the $\psi^{\rm pla}$ eigenstates entering into the evaluation of the  collision rate will carry an effective mixing angle $\vartheta_{\rm eff}$ rather than the vacuum one $\vartheta$, and can be strongly suppressed by $m_N^4/\Pi_\nu^2$. %

Note that in Eq.~\eqref{eq:constraint}, we have neglected the offshell broadening effect from the anti-commutator $\left\{\i\Sigma^{<}\gamma^0,\gamma^0\real S^{R}\right\}$ in the narrow-width approximation, where the real part of the retarded propagator $\real S^{R}$ represents the off-shell propagation. Following the same spirit of the narrow-width approximation, we have also neglected the collision rate, $[\gamma^0 \i S^<,\i\Sigma^>\gamma^0]_{ij}-[\gamma^0 \i S^>,i\Sigma^<\gamma^0]_{ij}$ in Eq.~\eqref{eq:constraint}, where $[X,Y]$ denotes the commutatator between $X$ and $Y$, and $\i\Sigma^{>,<}$ denotes one-loop amplitudes.  The collision rate in the constraint equation contributes as an imaginary pole at the onshell resonance.  This contribution becomes important near resonant production, such as resonant Dodelson-Widrow oscillation production~\cite{Shi:1998km}, but later we will show that such a resonance does not occur for collisional freeze-in production of sterile-neutrino DM in the parameter region of interest.

\paragraph{\textbf{Mixing-induced collisional freeze-in.}} We now show how the plasma induced rotation enters the kinetic equations. To this end, we employ  the resummed Wightman functions in the $\psi^{\rm pla}$ basis,
\begin{align}\label{eq:S<}
         \i S^<_{ij}(k)&=-2\pi P_L\slashed{k} \,\delta(k^2)\left[\theta(k_0)f_{ij}(k_0,t) \right. \nonumber \\ & \qquad\qquad \qquad \left. -\theta(-k_0) \left(\delta_{ij}-\bar{f}_{ij}(-k_0,t)\right)\right],
         \\
         \label{eq:S>}
         \i S^>_{ij}(k)&=-2\pi  P_L \slashed{k} \,\delta(k^2)\left[\theta(-k_0) \bar f_{ij}(-k_0,t)
         \right. \nonumber \\
         & \qquad\qquad \qquad   \left. 
         -\theta(k_0)\left(\delta_{ij}-f_{ij}(k_0,t)\right) \right],
\end{align}
where $P_L$ selects the left-handed $\psi$ states, $\delta_{ij}$ is the Kronecker delta and $\theta(k_0)$ the Heaviside step function in frequency. For simplicity, we omit the superscript "pla", and neglect the neutrino thermal mass and $N$ vacuum mass in the Dirac delta functions.
The matrices $f_{ij},\bar f_{ij}$ satisfy $f^*_{ij}=f_{ji},\bar f^*_{ij}=\bar f_{ji}$, where $f_{11}, \bar f_{11}$ are the distribution functions of neutrinos and anti-neutrinos, $f_{22}, \bar f_{22}$ those of $N$ particle  and $\bar N$ antiparticle, while  $f_{12},f_{21}$ are the off-diagonal correlations which we referred to above as flavour coherences. For freeze-in production, we may neglect both $f_{22}$ and $\bar f_{22}$ for the non-equilibrium particle $N$,
such that
\begin{align}\label{eq:S11=-fS22}
\theta(k_0)\i S^{<}_{11}(k)=-\theta(k_0)f_{11}^{\rm eq}(|\boldsymbol{k}|)\i S^{>}_{22}(k)\,,
\end{align}
where $f_{11}^{\rm eq}(|\boldsymbol{k}|)=(e^{|\boldsymbol{k}|/T}+1)^{-1}$ is the thermal distribution function.

The FCKB kinetic equations in the $\psi^{\rm pla}$ basis are constructed from $\i S^{<}_{ij}(k,t)$,
\begin{align}\label{eq:kinetic} \i\gamma^0\frac{\d}{\d t} \i S^{<}_{ij}=\left[\real(\Sigma^{R})\gamma^0,\gamma^0\i S^{<}\right]_{ij}-\frac{\i}{2}\left( \mathcal{C}_{ij}+\mathcal{C}^\dagger_{ij}\right),
\end{align}
where $\mathcal C_{ij}$ is the FCKB collision rate,
\begin{align}\label{eq:collision}
\mathcal C_{ij}\equiv\sum_k \left(\i\Sigma^>_{ik}\,\i S^<_{kj}-\i\Sigma^<_{ik}\,\i S^>_{kj}\right)\,.
\end{align}
In the interaction basis the only non-zero entry of the one-loop self-energy is $\i \Sigma^{ab}_{\nu}$  of the electron neutrino, where $a,b = \pm$ are indices of the Closed-Time-Path formalism~\cite{Chou:1984es,Calzetta:1986cq}, $\i\Sigma^{>}_{ij}\equiv \i \Sigma_{ij}^{-+},\i\Sigma^{<}_{ij}\equiv \i \Sigma_{ij}^{+-}$ and $\real\Sigma_\nu^R=\real\Sigma_\nu^{++}$.
 Rotating to the plasma basis with $\psi^{\rm pla} = U^\dag_{\rm eff} \psi^{\rm int}$, the amplitudes in~\eqref{eq:collision} read
\begin{align}\label{eq:Sigma_ij^ab}
\i\Sigma^{ab}_{ij}=U^*_{\text{eff},ei}\,U_{\text{eff},ej}\,\i\Sigma^{ab}_{\nu},
\end{align}
where $i(j)$ denotes the outgoing (incoming) state in the self-energy diagram. 

The FCKB kinetic equations for the occupation-number matrices $f_{ij}$ are obtained by taking the Dirac trace  of the Wightman functions  and integrating over positive frequencies in Eq.~\eqref{eq:kinetic}.
After applying Eq.~\eqref{eq:S11=-fS22}, Eq.~\eqref{eq:Sigma_ij^ab},  and the relation $\i S^>_{ij}=\i S^<_{ij}$ for $i\neq j$, we  can simplify the kinetic equations in the small effective mixing regime $\vartheta_{\rm eff}\ll 1$, such that 
\begin{align}\label{eq:df12/dt}
    \frac{{\d}f_{12}}{{\d}t}&=\i \int_0^\infty \frac{{ \d}k_0}{2\pi}\Tr\left[\Sigma^R_{\nu} \left(\i S_{12}^< -\vartheta_{\rm eff}\i S_{11}^<\right)\right],
      \\[0.2cm]
    \frac{{\d}f_{22}}{{\d}t}&=\i \int_0^\infty\frac{{ \d}k_0}{2\pi}\vartheta_{\rm eff}\Tr\left(\Sigma^R_{\nu}\i S_{12}^< - \Sigma^{R*}_{\nu}\i S_{21}^<\right)
    \nonumber \\[0.2cm]
    &\quad +\int_0^\infty \frac{{\d}k_0}{2\pi}2\vartheta^2_{\rm eff}\Tr\left(\imag(\Sigma^R_{\nu})\i S_{11}^<\right),\label{eq:df22/dt}
\end{align}
and $\d f_{21}/\d t$ can be obtained by the Hermitian condition $f_{21}=f_{12}^*$. The first line of Eq.~\eqref{eq:df22/dt} reduces to $\vartheta_{\rm eff}(\d f_{12}/\d t+\d f_{21}/\d t)=2\vartheta_{\rm eff}\real(\d f_{12}/\d t)$ in the narrow-width approximation. The contribution from flavour coherences to $f_{22}$ can be constructive or destructive, depending on the sign of $\real\Sigma^R_\nu$. In addition, given $\real(\d f_{12}/\d t)=\mathcal{O}(\vartheta_{\rm eff})$, the coherence contribution is formally of the same order in the effective mixing angle as the second line of Eq.~\eqref{eq:df22/dt}. Nevertheless, the flavour coherence is built on a timescale $\tau_{\rm coh}\sim[\vartheta_{\rm eff}\real(\Sigma^R_\nu)]^{-1}$, which is much shorter than the collisional freeze-in timescale $\tau_{\rm cfi}\sim[\vartheta_{\rm eff}^2\imag(\Sigma^R_\nu)]^{-1}$. Therefore, by the time collisional production becomes relevant, the kinetic evolution of the coherence has entered the  decaying regime, $f_{12}\ll 1$, and its net contribution to DM production is suppressed.

If this coherence contribution is neglected, the remaining term for $f_{22}$ reduces to a quantum version of the  Boltzmann collision rate. This follows from the optical theorem, whereby the one-loop self-energy $\imag \Sigma^R_\nu$ produces the tree-level thermalization rate after onshell cuts~\cite{Weldon:1983jn}. In particular, for sterile neutrinos with $\Sigma^R_\nu$ generated by the weak charged current, $\imag \Sigma^R_\nu$ corresponds to the decay rate of the weak gauge bosons, leading to a collisional freeze-in rate proportional to $\vartheta_{\rm eff}^2$. If the thermal correction $\Pi$ can be neglected, Eq.~\eqref{eq:theta_eff} indicates that $\vartheta_{\rm eff}\to \vartheta$ and Eq.~\eqref{eq:df22/dt} in the approximation of $f_{12},f_{21}\approx 0$ recovers the semiclassical Boltzmann equation with a $\vartheta^2$ collision rate, as employed in several contexts, e.g., Refs.~\cite{Coy:2021sse,Datta:2021elq,Li:2022bpp}. However, we will show below that this is not the case for sterile-neutrino DM in the keV-MeV mass range.

The above analysis also clarifies why simply adding vacuum and/or thermal masses to the Dirac delta functions in Eqs.~\eqref{eq:S<} and \eqref{eq:S>} is insufficient to capture the thermal effects. Such masses enter the distribution functions and $\imag \Sigma^R_\nu$ as subdominant additive corrections. In the freeze-in regime where $T\ll m_W$ with  $m_W$ the $W$-boson mass, the neutrino thermal mass and the DM vacuum mass change the rate only by relative corrections of  $\mathcal O (g_2^2T^4/m_W^4)$ and  $\mathcal O(m_N^2/m_W^2)$, respectively.
Moreover, including the vacuum mass through $\slashed{k}\to\slashed{k}+m_N\mathbbm1$ does not change the chirality-conserving self-energy amplitudes, because the Dirac trace does not select the chirality-flipping $m_N\mathbbm1$ term in Eq.~\eqref{eq:collision}. Thermal masses likewise modify the phase-space integration, as expected in Boltzmann equations. The all-important multiplicative suppression, however,  originates from the rotation between the $\psi^{\rm vac}$ and $\psi^{\rm pla}$ bases. This rotation is determined by the constraint equation and is not visible in a Boltzmann treatment that only modifies the dispersion relations.

\paragraph{\textbf{Sterile-neutrino DM from collisional freeze-in}}
We now apply the above result to the collisional freeze-in production  of relativistic sterile-neutrino DM from the SM plasma, which is dominated by the $W$-boson decay~\cite{Lello:2016rvl,Datta:2021elq}. The one-flavor mixing case considered here can be generalized to multi-flavor mixing, and the SM plasma suppression still applies.  Nevertheless, the following conclusion should be revised properly when additional nonthermal states are introduced, as some channels beyond the standard, minimal thermal production may open for sterile neutrino DM, such as resonant oscillation with large asymmetry~\cite{Shi:1998km,Asaka:2005an} or heavy particle decay~\cite{Kusenko:2006rh,Petraki:2007gq,Abada:2025gvc}. We neglect the contribution from flavour coherence, the first integral of~\eqref{eq:df22/dt}, as it is suppressed on the collisional time scale.
In the relevant low-temperature regime $z \equiv m_W/T \gg 1$, we include both the charged and neutral weak currents to evaluate $\Pi_\nu$, giving $\Pi_\nu \approx-
{8g_2^2(2+\cos^2(\theta_W))|\k|^2T^4}/({\pi^2m_W^4})$,
and hence from Eq.~\eqref{eq:theta_eff} we obtain a suppression factor $m_N^2/m_W^2$ for $\vartheta_{\rm eff}$,
\begin{align}\label{eq:vartheta_eff}
    \vartheta_{\rm eff}
    \approx 2 \times 10^{-9}\left(\frac{3}{x}\right)^2 
    \left(\frac{z}{10}\right)^6\left(\frac{m_N}{10\ {\rm keV}}\right)^2 \vartheta\,,
\end{align}
where $x\equiv|\k|/T\sim z$ for nonrelativistic $W$-boson decay.
As we show in the End Matter, freeze-in peaks at $z\simeq 10$ and the $z^{-1}$ expansion in~\eqref{eq:vartheta_eff} is well justified. The associated temperature is ``high'', well above the Dodelson-Widrow production of $T\sim 100-200$~MeV 
but ``low'' compared to the electroweak symmetry breaking epoch. Compared to the claim in~\cite{Datta:2021elq}, we find a net reduction of the relic abundance by 20 orders of magnitude at $m_N = 10\ {\rm keV}$; see Fig.~\ref{fig:YmN}. Our result therefore aligns with the one derived in Ref.~\cite{Lello:2016rvl}.
We conclude that the collisional freeze-in production rate of sterile-neutrino DM from gauge-boson decay is not controlled  by the vacuum quantity $\vartheta^2$, but instead by~$\vartheta_{\rm eff}^2$.  For  production of keV-MeV scale sterile neutrinos at $z\simeq 10$, it is easy to verify that the resonance point $\Pi_\nu=m_N^2$ cannot be met because $\Pi_\nu<0$. At higher temperatures when $\Pi_\nu$ becomes positive, the enhancement near the resonant point will be suppressed by the offshell broadening effect we neglected in Eq.~\eqref{eq:constraint}. Physically, including  the broadening effect corresponds to the  finite damping rate discussed in Ref.~\cite{Lello:2016rvl}, where the enhancement  effect was found to be numerically small.

\begin{figure}[t]
	\centering
\includegraphics[width=\columnwidth]{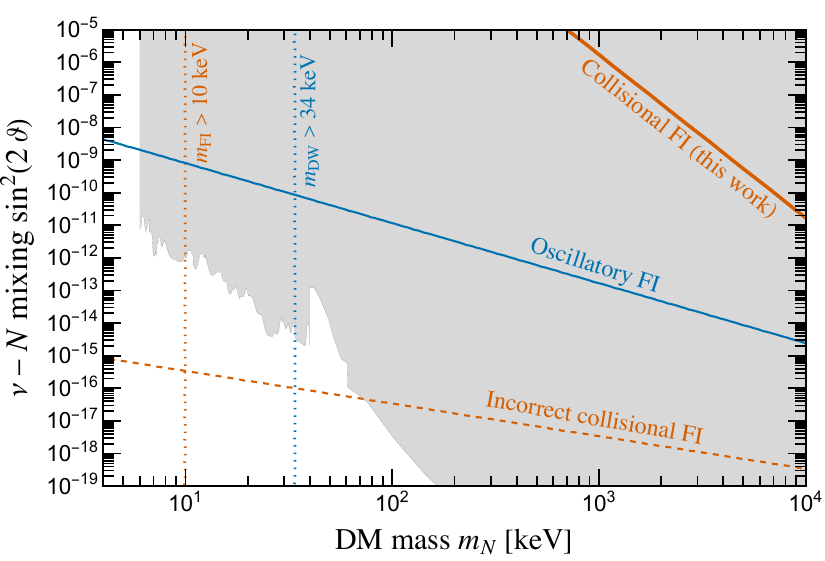}
	\caption{\label{fig:nuDM} Sterile-neutrino DM production from  collisional freeze-in (FI), in comparison with the  nonresonant Dodelson-Widrow oscillatory freeze-in. 
    The incorrect collisional FI is from the semiclassical Boltzmann kinetics.  The vertical dotted lines represent lower DM mass bounds from freeze-in ($m_{\rm FI}$) and nonresonant oscillation ($m_{\rm DW}$) production, respectively, and excluded regions are shown in gray.
    }
\end{figure}

We show in Fig.~\ref{fig:nuDM} the production rates that yield the observed DM relic density, where we apply the semi-analytic fit of the mass-mixing angle relation for the nonresonant Dodelson-Widrow oscillatory freeze-in~\cite{Asaka:2006rw,Asaka:2006nq}.
The freeze-in production rate obtained at $\mathcal{O}(\vartheta^2)$ overestimates the relic density by many orders of magnitude for keV-MeV sterile-neutrino DM. From Fig.~\ref{fig:nuDM}, we see that  the $\vartheta_{\rm eff}^2$ collisional freeze-in rate requires a much larger vacuum mixing angle than the non-resonant Dodelson-Widrow production for $m_N<10$~MeV. Both production rates are computed from effective mixing angle $\vartheta_{\rm eff}$ and only differ in the collisional SM rates that multiply them. Interestingly, we noticed that when DM becomes heavier than about 230~MeV, the collisional freeze-in production can become dominant over  the nonresonant Dodelson-Widrow oscillatory freeze-in. At still larger $m_N$, the $\vartheta_{\rm eff}^2$ collisional freeze-in rate approaches the $\vartheta_{}^2$ rate and the FCKB kinetics reproduces the Boltzmann kinetics, as shown in Fig.~\ref{fig:YmN}. While a sterile neutrino much heavier than the MeV scale generally does not play the role of a stable DM candidate, collisional freeze-in  opens a production channel for heavier nonthermal particles that couple to a thermal plasma via small mixing.  This is of particular interest for cosmologically long-lived particles. 

Structure formation excludes too light sterile neutrinos.  Lyman-$\alpha$ forest data require $m_N > m_{\rm FI}\simeq 10$~keV for DM produced by freeze-in from decays~\cite{Irsic:2017ixq,Ballesteros:2020adh}, 
and $m_N > m_{\rm DW}>34$~keV  for
 nonresonant Dodelson-Widrow production~\cite{Irsic:2017ixq,Baur:2017stq}. When combined with the bounds from X-ray and gamma-ray observations~\cite{Boyarsky:2007ge,Roach:2022lgo,Foster:2021ngm,Calore:2022pks,Krivonos:2024yvm}, shown by the gray shaded region in Fig.~\ref{fig:nuDM}, both the production from the W-boson decay and the nonresonant oscillation that yield the observed relic density are ruled out for keV-MeV scale sterile neutrinos. This conclusion does not change when the coherence contributions are included:  we have  checked numerically that the first line of Eq.~\eqref{eq:df22/dt} is small relative to the $\vartheta_{\rm eff}^2$ rate for the parameter space shown in Fig.~\ref{fig:nuDM}.

\paragraph{\textbf{Conclusions}} We have developed a quantum flavour-covariant formalism tailored to collisional freeze-in production of light nonthermal particles through mixing portals, demonstrating strong suppression effects from thermal plasma and the asymptotically approximate equivalence between the quantum FCKB and semiclassical Boltzmann kinetics in terms of the vacuum mass. Applying the FCKB formulation to sterile-neutrino DM resolves the conflicting conclusions about the efficiency and viability of collisional freeze-in from gauge interactions, where the semiclassical Boltzmann treatment overestimates the keV-MeV DM relic density by a factor of $10^{12}-10^{24}$. The technique has
broad implications in astrophysics and the early Universe, allowing to properly evaluate collisional freeze-in production of nonthermal particles through mixing-induced interactions in a hot plasma.

\paragraph{\textbf{Acknowledgements}} We would like to thank Maxim Pospelov for helpful discussions, and appreciate  comments from Apostolos Pilaftsis and Arghyajit Datta. Cofunded by the European Union (ERC, NLO-DM, 101044443). This work was also supported by the Research Network Quantum Aspects of Spacetime (TURIS).

\bibliographystyle{JHEP}
\bibliography{Refs}

\StartEndMatter

\paragraph{\textbf{Rotation from vacuum to thermal basis.}} Here we show how the inclusion of thermal corrections $\real\Sigma_\nu^R$ introduces a further rotation from the vacuum basis $\psi^{\rm vac}$ to the plasma basis $\psi^{\rm pla}$ in the FCKB formalism. We can  rewrite the FCKB constraint equation~\eqref{eq:constraint} in the vacuum basis as
\begin{align}\label{eq:FCKB-constraint}
\left[\slashed k-\hat{\mathbb{M}}-\operatorname{Re} \Sigma^R\right] \mathrm{i} S^{<}  +\rm h.c. =0 .
\end{align}
In the vacuum-mass basis $\real\Sigma^R_{ij}=U^*_{ei}U_{ej}\real\Sigma^R_\nu$, see~\eqref{eq:Sigma_ij^ab} with $U_{\rm eff}\to U$. The displayed equation is the left-acting Kadanoff-Baym equation $(\slashed{k}-\mathbb{\hat M}-\real\Sigma^R)\,\i S^<=0$ and the Hermitian conjugate (h.c.) part is the right-acting one, up to $\i \gamma^0 \partial_t/2$ which cancels in the sum.
In the quasiparticle approximation both lead to the same onshell condition, and it is therefore sufficient to consider the first term in~\eqref{eq:FCKB-constraint}.

Due to the chiral structure of the weak interactions, $\real\Sigma^R_\nu=P_R\,\real\Sigma^R_\nu P_L$ and in a homogeneous, isotropic plasma we may decompose $\real\Sigma_\nu^R$  as~\cite{Weldon:1982bn,Li:2023ewv}
\begin{align}
\label{eq:weldondecomp}
    \real\Sigma_\nu^R=-a\slashed{k}P_L-b  \slashed{u}P_L\,,
\end{align}
where $a,b$ are coefficients, and $u^\mu=(1,\boldsymbol{0})$ is the four-velocity in the  plasma rest frame.
In vacuum, with $\real\Sigma^R=0$, the first term in~\eqref{eq:FCKB-constraint} reduces to $(\slashed{k}-\mathbb{\hat M})\i S^{<}=0$. It has non-trivial solutions when the Dirac determinant
$\text{det}_{\rm s}(\slashed{k}-\hat m_i)=(k^2-\hat m_i^2)^2 = 0$ for some~$i$, 
giving the usual quasiparticle solution 
$\i S^<_{ii}\propto(\slashed{k}+\hat m_i)\,\delta(k^2-\hat m_i^2)$ for the diagonal elements.

With thermal corrections, we multiply the first term of~\eqref{eq:FCKB-constraint} from the left by $\slashed k+\mathbb{\hat M}$, which yields,
\begin{align}\label{eq:FCKB-constraint-2}
\left[k^2-\mathbb{\hat M}^2-\slashed k\,\real\Sigma^R-\mathbb{\hat M}\,\real\Sigma^R\right]\i S^<=0\,.
\end{align}
The last term flips chirality and is suppressed by $\hat m_i/|\k|$ relative to the third term. To evaluate the latter we use~\eqref{eq:weldondecomp}, $\slashed k\real\Sigma^R_\nu=-\left(ak^2+bk_0+b\,\gamma^0\k\cdot\boldsymbol\gamma\right)P_L$, and, in the chiral representation with $\mathbf k$ along the $z$-axis,  $\gamma^0\k\cdot\boldsymbol\gamma P_L =-|\mathbf k|\,\diag(\sigma^3,0)$. We see that $\slashed k \real\Sigma^R_\nu$ is diagonal with the two non-zero entries acting on the left-chiral components with spin parallel and anti-parallel  to $\mathbf k$, respectively. At $k_0 = \pm |\mathbf{k}|$ the propagating left-chiral component is the one with spin antiparallel (parallel) to  $\mathbf{k}$, the neutrino (antineutrino), for which  $\slashed k \real\Sigma^R_\nu\simeq-(ak^2+2bk_0)$; the other entry is $\mathcal O(k^2)$ and leaves the vacuum pole unchanged.  Up to  $\mathcal O(\tilde m_i^2/|\k|^2)$ we may therefore replace $\slashed k \real\Sigma^R\to\Pi$, with $\Pi_{ij}=U^*_{ei}U_{ej}\Pi_\nu$ and 
\begin{align}\label{eq:Pi-EM}
\Pi_\nu\equiv\Tr_{\rm s}\big[\slashed k\,\real\Sigma^R_\nu\big]=-2\left(ak^2+bk_0\right)\simeq - 2 b k_0.
\end{align}
Equation~\eqref{eq:FCKB-constraint-2} then becomes $\big[k^2-(\mathbb{\hat M}^2+\Pi)\big]\i S^<\simeq0$. For a single massless flavour this is $k^2=\Pi_\nu$, the in-medium dispersion relation $k_0\simeq|\k|-b$~\cite{Weldon:1982bn,Notzold:1987ik}. It can be shown, that the same pole condition follows from $\det(\slashed k-\mathbb{\hat M}-\real\Sigma^R)=0$.

Neither the diagonal elements of the free Wightman function, $\i S^<_{ii}\propto \delta(k^2-\hat m_i^2)$, nor the same elements with shifted poles, $\i S^<_{ii}\propto\delta(k^2-\tilde m_i^2)$, solves~\eqref{eq:FCKB-constraint-2}, because $\mathbb{\hat M}^2+\Pi$ is not diagonal in the vacuum basis: for fixed $j$, the column $(\i S^<_{1j},\i S^<_{2j} )$ must be an eigenvector of $\mathbb{\hat M}^2+\Pi$ with $k^2$ the corresponding eigenvalue~$\tilde m_i^2$.
Instead, one performs the further rotation $\psi^{\rm pla}=V\psi^{\rm vac}$, $\i S^{<,\rm pla}=V\,\i S^{<,\rm vac}V^\dagger$, with $V$ defined by Eq.~\eqref{eq:diagonalization}. In the $\psi^{\rm pla}$ basis, \eqref{eq:FCKB-constraint-2} reduces to $(k^2-\tilde m_i^2)\,\i S^{<,\rm pla}_{ij}\simeq0$, so that $\i S^{<,\rm pla}_{ii}\propto\delta(k^2-\tilde m_i^2)$ is the quasiparticle solution in the plasma.
Under this rotation, $(V\real\Sigma^RV^\dagger)_{ij}=(UV^\dagger)^*_{ei}(UV^\dagger)_{ej}\real\Sigma^R_\nu$, which is Eq.~\eqref{eq:Sigma_ij^ab} with $U_{\rm eff}=UV^\dagger$.  In the interaction basis $U\Pi U^\dagger=\diag(\Pi_\nu,0)$, so that for $\hat m_1=0$
\begin{align}\label{eq:Hint}
U(\mathbb{\hat M}^2+\Pi)U^\dagger=\begin{pmatrix} m_N^2s^2_\vartheta+\Pi_\nu & m_N^2s_\vartheta c_\vartheta\\ m_N^2s_\vartheta c_\vartheta & m_N^2 c^2_\vartheta\end{pmatrix},
\end{align}
with $s_\vartheta\equiv \sin \vartheta, c_\vartheta\equiv \cos\vartheta$,
and the diagonalization by $U_{\rm eff}$ gives Eq.~\eqref{eq:theta_eff}.

It remains to evaluate $\Pi_\nu$ in Eq.~\eqref{eq:Pi-EM} in the regime relevant for freeze-in, $z=m_W/T\gg1$. With $\real \Sigma_\nu^R =\real(\Sigma_\nu^{++})$, where  $++$ is the  index in the Closed-Time-Path space denoting the time-ordered quantity, we can calculate the real part of the time-ordered one-loop self-energy diagrams.  The self-energy receives charged- and neutral-current contributions, $ \real \Sigma_\nu^R = \real \Sigma_W^R + \real \Sigma_Z^R $.
For the charged weak current, we can write down the amplitude  as 
\begin{align}
  \Sigma_W^{++}=\frac{g_2^2}{2}\int \frac{\d^4 k_e}{(2\pi)^4}&\frac{\d^4 k_W}{(2\pi)^4}(2\pi)^4 \delta^4(k_W+k-k_e)
  \nonumber \\
  &\times \gamma^\mu P_L\i S_e^{++}\gamma^\nu P_L \i G_{\mu\nu}^{++}\,,
\end{align}
with time-ordered propagators 
\begin{align}
    \i S_e^{++}(k)&=\frac{\i (\slashed{k}+m_e)}{k^2-m_e^2+\i \epsilon}
    \nonumber \\
    &-2\pi (\slashed{k}+m_e)\delta(k^2-m_e^2)f_e(|k_0|)\,,
    \\
\i G_{\mu\nu}^{++}(k)&=\frac{-\i \eta_{\mu\nu} }{k^2-m_W^2+\i \epsilon}
\nonumber \\ 
&+2\pi \eta_{\mu\nu}\delta(k^2-m_W^2)f_W(|k_0|)\,.
\end{align}
Neglecting the electron and DM masses as well as the $W$ distribution $f_W$, which is suppressed by $e^{-z}$, and using Boltzmann-statistics, 
the Dirac trace 
 gives
the $W$-boson contribution,
\begin{align}
   \Pi_W= \Tr_{\rm s}\big[\slashed k \real\Sigma^R_W\big]  \approx\frac{-16g_2^2|\k|^2T^4}{\pi^2m_W^4}\,,
\end{align}
where $k_0\approx |\k|$ was used, up to subdominant corrections at $\mathcal{O}(m_e^2/m_W^2), \mathcal{O}({\tilde m}_i^2/m_W^2)$.
By inspection of~\eqref{eq:Pi-EM}, this corresponds to $b=-\Pi_W/(2|\k|)=8g_2^2|\k|T^4/(\pi^2m_W^4)$ in Eq.~\eqref{eq:weldondecomp}, i.e., the neutrino potential $\mathcal{V}_{ee} = -b$ of Ref.~\cite{Notzold:1987ik}. 
The neutral-current contribution follows by $g_2/\sqrt2\to g_2/(2\cos\theta_W)$, $m_W\to m_Z=m_W/\cos\theta_W$ and $f_e\to f_\nu$; for a common neutrino and electron temperature, this gives $\Pi_Z\approx\cos^2\theta_W\,\Pi_W/2$.

It is worth making a comparison with the effective potential derived in the Dodelson-Widrow oscillatory freeze-in, where the effective mixing angle is obtained by diagonalizing  the effective Hamiltonian entering the coherent oscillation source in the density-matrix formalism~\cite{Sigl:1993ctk,Akhmedov:1998qx},
$    \mathcal{H}=\mathcal{V}+U\mathbb{\hat M}^2 U^\dagger/(2|\k|)$,
where the in-medium potential $\mathcal V$ has the non-zero entry $\mathcal V_{ee}$; the factor $1/(2|\k|)$ stems from the relativistic expansion of $\sqrt{|\k|^2+m^2}$.
Since $2|\k|\mathcal H=U\mathbb{\hat M}^2U^\dagger+\diag(2|\k|\mathcal V_{ee},0)$ coincides with Eq.~\eqref{eq:Hint} for $\Pi_\nu=2|\k|\mathcal V_{ee}$, the mixing angles in oscillatory and collisional freeze-in are identical, Eq.~\eqref{eq:theta_eff}.

\paragraph{\textbf{Freeze-in temperature and yield.}} The production rate for collisional freeze-in from $W$ gauge boson decay is given by momentum-integrating the second integral of the right-hand-side of~\eqref{eq:df22/dt}
\begin{align}
     \gamma_{\rm prod} &= 2\int \frac{\d^4 k}{(2\pi)^4}\theta(k_0)\vartheta^2_{\rm eff}\Tr\left(\imag(\Sigma^R_{\nu})\i S_{11}^<\right)
    \nonumber\\
     &=-\int \frac{\d^3 \k}{(2\pi)^3 |\k|}\vartheta^2_{\rm eff} f_{11}^{\rm eq}(|\k|)\Tr\left(\imag(\Sigma^R_{\nu})\slashed{k}\right).\label{eq:gamma-pro}
\end{align}
Here, $\imag \Sigma^R_{\nu} =\left(\i\Sigma_\nu^<(k)-\i\Sigma_\nu^>(k)\right)/2$ is the imaginary part of the retarded self-energy amplitude, where
\begin{align}\label{eq:TrImR-fin} 
   & f_{11}^{\rm eq}(|\k|)\Tr[\imag(\Sigma_\nu^R)\slashed{k}]
 \\
    &=-\frac{g_2^2m_W^2}{16\pi|\k|}\int_{m_W^2/(4|\k|)}^\infty \d E_ef_W(E_e+|\k|)\left(1-f_e(E_e)\right), \nonumber 
\end{align}
with $f_W, f_e$ the distribution functions of $W$ boson and electron, respectively. Using~\eqref{eq:TrImR-fin}  together with $|\Pi_\nu|\gg m_N^2$ for which $\vartheta_{\rm eff}^2 \simeq \vartheta^2 m_N^4/\Pi_\nu^2 $, the collision term~\eqref{eq:gamma-pro} can be integrated in the Maxwell-Boltzmann approximation. Further taking $1-f_e \simeq 1$,  we arrive at 
\begin{align}
    \label{eq:gamma-analyt}
    \gamma_{\rm prod} \simeq \frac{3\pi^2m_N^4 z^7 K_3(z)  }{8g_2^4(2+\cos^2\theta_W)^2 m_W}\times \vartheta^2 \Gamma_{W\to e\nu} \,.
\end{align}
Here, we have introduced the vacuum $W$-boson decay width to a lepton-neutrino pair, $\Gamma_{W\to e\nu} = g_2^2 m_W/(48\pi) \simeq 0.227$~GeV; $K_3(x)$ is the modified Bessel function of the second kind.

\begin{figure}[b!]
	\centering
\includegraphics[width=\columnwidth]{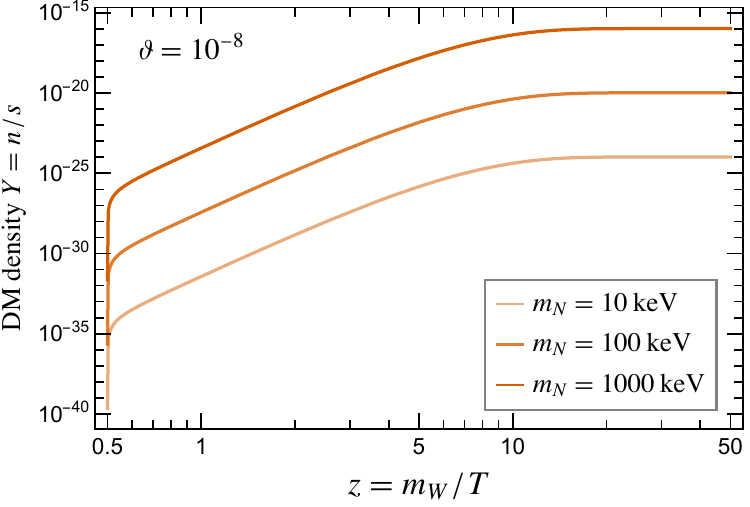}
	\caption{\label{fig:CFIpeak}The evolution of DM yield $Y(z)=n/s$ at $\vartheta=10^{-8}$. The initial condition with $Y=0$ is set at $z = 0.5$, corresponding to the electroweak crossover temperature $T_c\approx 160$~GeV when the mixing angle $\theta$ is generated.}
\end{figure}

Denoting the sterile neutrino number density $n$ normalized to the entropy density $s = (2\pi^2/45) g_{\star s}T^3$ by~$Y\equiv n/s$, the abundance evolution is obtained by integrating
\begin{align}
\label{eq:yield}
   \frac{\d Y}{\d z} =  \frac{\gamma_{\rm prod}}{sHz},
\end{align}
 where 
 \begin{align}
    H = \frac{\sqrt{g_\star \pi^2/90} T^2}{M_P}\,,
 \end{align}
 is the Hubble rate during the radiation dominated epoch; $M_P \simeq 2.4 \times 10^{18}$~GeV is the reduced Planck mass. With the collision term~\eqref{eq:gamma-analyt}, it is noticeable that the scale of the production rate is determined by the sterile-neutrino mass $\gamma_{\rm prod}\propto m_N^4$, and the integrand $\d Y$ becomes proportional to $z^{11} K_3(z)$ which peaks at $z\simeq 10$ corresponding to a characteristic freeze-in temperature of~8~GeV; see also Fig.~\ref{fig:CFIpeak}. The final $z$-integral may to good approximation be preformed over an unrestricted range. Using the characteristic values $g_\star \simeq g_{\star s} \simeq 86$, we obtain
 \begin{align}
 Y(z\to \infty) \simeq 10^{-8} \vartheta^2 \left( \frac{m_N}{10~{\rm keV}} \right)^4\,.
 \end{align}
 We may convert this into a relic density parameter  using $\Omega_N = {m_N s_0 h^2 Y}/\rho_c $ with $s_0 h^2/\rho_c \simeq 2.75 \times 10^8\ {\rm GeV^{-1}}$~\cite{ParticleDataGroup:2026aaa},
\begin{align}
    \Omega_N h^2 \simeq  2.8\times 10^{-5}\vartheta^2 \left(\frac{m_N}{10~\mathrm{keV}}\right)^5. 
\end{align}

\end{document}